\documentclass[prd,twocolumn,aps,preprintnumbers,amsmath,amssymb,footinbib,superscriptaddress,showpacs]{revtex4}

\usepackage{color}
\usepackage{graphicx}
\usepackage{tabularx}
\usepackage{amsmath}
\usepackage{amssymb}

\newcommand{\ssection}[1]{{\em #1.\ }}

\begin{document}

\title{Enhancement of Dark Matter Annihilation via Breit-Wigner Resonance}
\author{Wan-Lei Guo and Yue-Liang Wu}
\affiliation{ Kavli Institute for Theoretical Physics China, \\
Key Laboratory of Frontiers in Theoretical Physics, \\ Institute of
Theoretical Physics, Chinese Academy of Science, Beijing 100190,
China}
\date{\today}

\begin{abstract}
The Breit-Wigner enhancement of the thermally averaged annihilation
cross section $\langle \sigma v \rangle$ is shown to provide a large
boost factor when the dark matter annihilation process nears a
narrow resonance. We explicitly demonstrate the evolution behavior
of the Breit-Wigner enhanced $\langle \sigma v \rangle$ as the
function of universe temperature for both the physical and
unphysical pole cases. It is found that both of the cases can lead a
large enough boost factor to explain the recent PAMELA, ATIC, and
PPB-BETS anomalies. We also calculate the coupling of the
annihilation process, which is useful for an appropriate model
building to give the desired dark matter relic density.
\end{abstract}

\pacs{95.35.+d, 98.70.Sa}

\maketitle

\ssection{Introduction} The existence of dark matter is by now well
confirmed \cite{DM}. However, there is no candidate for the dark
matter in the standard model.  Understanding the nature of dark
matter is one of the most challenging problems in particle physics
and cosmology. The recent cosmological observations have established
the concordance cosmological model where the present energy density
consists of about 73\% dark energy, 23\% dark matter, and  4\% atoms
\cite{WMAP}. Currently, many dark matter search experiments are
under way. These experiments can be classified as the direct dark
matter searches and the indirect dark matter searches. The direct
dark matter detection experiments may observe the elastic scattering
of dark matter particles with nuclei. The indirect dark matter
searches are designed to detect the dark matter annihilation
productions, which include neutrinos, gamma rays, electrons,
positrons, protons and antiprotons. In addition, the CERN LHC
searches are complementary to the direct and indirect dark matter
detection experiments.


Recently, the indirect dark matter detection experiment PAMELA
\cite{PAMELA} reported an excess in the positron fraction from 10 to
100 GeV, but showed no excess for the antiproton data. The ATIC
\cite{ATIC} and PPB-BETS \cite{PPB} balloon experiments have also
seen the excess in the $e^+ + e^-$ energy spectrum between 300 and
800 GeV. It is a natural idea that the dark matter annihilation can
account for the PAMELA, ATIC, and PPB-BETS anomalies. However, the
thermally averaged annihilation cross section $\langle \sigma v
\rangle$ obtained from the observed relic density is far smaller
than the required value from the PAMELA data. Therefore, one must
resort to the large boost factor (about $100-1000$) to explain the
large positron flux. Current analysis on the clumpiness of dark
matter structures indicates that the  most probable boost factor
should be less than $10-20$ \cite{Structure}. Considering the
difficulty to yield a large boost factor, many authors investigate
the decaying dark matter \cite{Decay}. However, the PAMELA data
require the lifetime of dark matter to be of the order of $10^{26}$
s. An alternative opinion is the nonperturbative Sommerfeld
enhancement, which may provide a large boost factor as the weak
force enhances the annihilation cross sections in the galactic halo
\cite{Sommerfeld}.


The thermally averaged annihilation cross section $\langle \sigma v
\rangle$ is a key quantity in the determination of the cosmic relic
abundances of dark matter. On the other hand, $\langle \sigma v
\rangle$ also determine the dark matter annihilation rate in the
galactic halo. The only difference among the above two cases is the
temperature $T$. For the relic density, $\langle \sigma v \rangle$
is usually evaluated at the freeze-out temperature $x \equiv m/T
\approx 20$ (the averaged velocity $v \approx \sqrt{3/x}$), where
$m$ is the dark matter mass. The dark matter annihilation in the
galactic halo occurs at $x \approx 3 \times 10^6$ ($v \approx
10^{-3}$). For nonrelativistic gases, $\langle \sigma v \rangle$ can
usually be expanded in powers of $x$, $\langle \sigma v \rangle
\propto x^{-k}$ \cite{KOLB}. For the $s$-wave annihilation ($k=0$),
$\langle \sigma v \rangle$ is a constant, which is independent of
the temperature of the Universe. For the $p$-wave annihilation
($k=1$), $\langle \sigma v \rangle$ will be decreased as the
Universe evolution. Clearly, only if $k < 0$, $\langle \sigma v
\rangle$ could be enhanced at the lower temperature. In such a case,
one may obtain a large boost factor to explain the PAMELA, ATIC, and
PPB-BETS anomalies. It is interesting to notice that when
considering the annihilation cross section at a narrow resonance, we
can derive a negative number for $k$, which then indicates a
Breit-Wigner enhancement mechanism. Recently, such an enhancement
has explicitly been analyzed in Ref. \cite{Murayama} (for the
previous discussions, see  Ref. \cite{Prev}). In the past, many
authors have studied the dark matter annihilation near a resonance
\cite{Griest, APP}.


In this paper, we try to further give a comprehensive analysis on
such a Breit-Wigner enhancement. Instead of using the center of mass
frame, we work in the cosmic comoving frame and adopt the usual
single-integral formula to calculate  $\langle \sigma v \rangle$.
Except for checking the unphysical pole case, we will pay attention
to the investigation for the physical pole case in which the cross
section $\langle \sigma v \rangle$ is found to have a maximum. In
both cases, $\langle \sigma v \rangle$ will approach a constant as
the Universe evolution. In terms of the observed dark matter
abundance, we calculate the coupling of the annihilation process for
the whole resonance parameter space. Hence, we derive the exact
boost factor and find that both cases can lead a large enough boost
factor to account for the PAMELA, ATIC, and PPB-BETS results.



\ssection{Breit-Wigner enhancement} The PAMELA experiment observing
no excess for the antiproton data indicates that dark matter will
dominantly annihilate into the leptonic final states. In fact, the
dark matter may first annihilate into some particles, such as the
Higgs triplets in the left-right symmetric model \cite{LR}, and then
these particles decay into the charged leptons. For the purpose of
this paper, we simply consider that two dark matter particles
directly annihilate into a pair of charged leptons via $S$ channel
Higgs boson exchanging. Since the Breit-Wigner enhancement requires
a narrow resonance, we follow Ref. \cite{Murayama} to introduce an
auxiliary parameter $\delta$ ($|\delta| \ll 1$) to express the
intermediate particle mass $M$
\begin{eqnarray}
M^2 = 4 m^2 (1- \delta) \;.
\end{eqnarray}
For the $\delta < 0$ case, one may obtain a physical pole. In this
case, the exchanging particle may decay into both initial states and
final states. For a given decay width $\Gamma$, one may write the
following annihilation cross section
\begin{eqnarray}
4 E_1 E_2 \sigma v = \frac{32 \pi}{\sqrt{1 - \frac{4 m^2}{M^2}}}
\frac{s}{M^2} \frac{M^2 \Gamma^2}{(s-M^2)^2 + M^2 \Gamma^2} B_i B_f,
\label{neg}
\end{eqnarray}
where $B_i$ and $B_f$ are the branching fractions of the resonance
into the initial and final channels, respectively. Because of $B_i +
B_f =1$, one can directly obtain $B_i B_f \leq 0.25$. Here we have
neglected the masses of final leptons. The parameter $s$ is defined
by $s \equiv (p_1 + p_2)^2$, where $p_1$ and $p_2$ are the
four-momenta of initial dark matter particles. For the $\delta > 0$
case, we have an unphysical pole. In this case, the intermediate
particle can not decay into the initial dark matter particles.
Therefore, we introduce a vertex $\alpha$ for the trilinear coupling
among two dark matter particles and the exchanging particle to
express the annihilation cross section
\begin{eqnarray}
4 E_1 E_2 \sigma v = 2 \alpha^2 \frac{s}{M^2} \frac{M
\Gamma}{(s-M^2)^2 + M^2 \Gamma^2} \;. \label{pos}
\end{eqnarray}

For the thermally averaged annihilation cross section $\langle
\sigma v \rangle$,  we adopt the usual single-integral formula
\begin{eqnarray}
\langle \sigma v \rangle = \frac{1}{n_{EQ}^2} \frac{m}{64 \pi^4 x}
\int_{4 m^2}^{\infty} \hat{\sigma}(s) \sqrt{s} K_1(\frac{x
\sqrt{s}}{m}) d s \;, \label{cross}
\end{eqnarray}
with
\begin{eqnarray}
n_{EQ} & = & \frac{g_i}{2 \pi^2} \frac{m^3}{x} K_2(x) \; ; \label{n} \\
\hat{\sigma}(s) & = & 4 E_1 E_2 \sigma v \; g_i^2 \; \sqrt{1-\frac{4
m^2}{s}} \; , \label{sigmahat}
\end{eqnarray}
where $K_1(x)$ and $K_2(x)$ are the modified Bessel functions. $g_i
=1$ is the internal degrees of freedom of dark matter particle.
Since $v$ of $\langle \sigma v \rangle$ is the M{\o}ller velocity,
we work in the cosmic comoving frame. If one takes the center of
mass frame  \cite{Murayama}, $\langle \sigma v \rangle$ should be
multiplied by a factor $(1+K_1^2(x)/K_2^2(x))/2$ \cite{APP}.

\begin{figure}[t]\begin{center}
\includegraphics[scale=0.45]{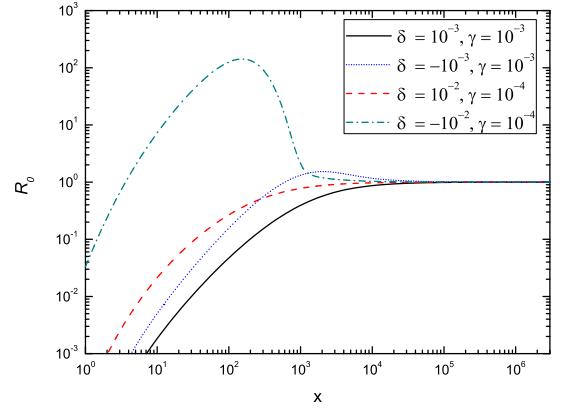}
\end{center}
\vspace{-0.5cm} \caption{Numerical illustration of the Breit-Wigner
enhanced $\langle \sigma v \rangle$ as a function of  $x$.  The
parameter $R_0$ is defined as $R_0 \equiv \langle \sigma v \rangle /
\langle \sigma v \rangle_0$, with $\langle \sigma v \rangle_{0}$
denoting the thermally averaged annihilation cross section at $T =
0$. } \label{f1}\end{figure}

\begin{figure}[t]\begin{center}
\includegraphics[scale=0.45]{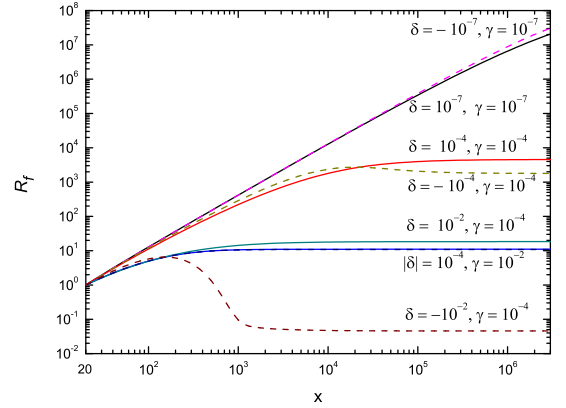}
\end{center}
\vspace{-0.5cm} \caption{Numerical illustration of the Breit-Wigner
enhanced $\langle \sigma v \rangle$ as a function of $x$. The
parameter $R_f$ is defined as $R_f \equiv \langle \sigma v \rangle /
\langle \sigma v \rangle_{x = 20}$, with $\langle \sigma v
\rangle_{x = 20}$ denoting the thermally averaged annihilation cross
section at the usual ``freeze-out"  time $x = 20$. }
\label{f2}\end{figure}

For an analytical illustration, we redefine
\begin{eqnarray}
s \equiv 4m^2 (1 + z) \;.
\end{eqnarray}
Then the integration region of Eq. (\ref{cross}) becomes $0 \leq z <
\infty$.  The annihilation cross sections $4 E_1 E_2 \sigma v$ in
Eqs. (\ref{neg}) and (\ref{pos}) can be rewritten as
\begin{eqnarray}
4 E_1 E_2 \sigma v \propto \frac{1 + z}{(z + \delta)^2 + \gamma^2}
\;,
\end{eqnarray}
where $\gamma$ is given by
\begin{eqnarray}
\gamma \equiv \Gamma/M \;.
\end{eqnarray}
For $x \geq 20$ and $|\delta|, \gamma \leq 0.1$,  $\langle \sigma v
\rangle$ in Eq. (\ref {cross}) can approximately be given by
\begin{eqnarray}
\langle \sigma v \rangle  \propto x^{\frac{3}{2}} \int_0^{z_{eff}}
\frac{ e^{-x z} \sqrt{z}}{(z +\delta)^2 +\gamma^2} \; d z \;.
\label{10}
\end{eqnarray}
It is worthwhile to stress that the integration result of Eq.
(\ref{10}) is insensitive to $x$ when $z$ is negligible in $(z
+\delta)^2 +\gamma^2$. In the $\delta > 0$ case, the effective
integration upper bound is $z_{eff} \sim 4/x$, thus $\langle \sigma
v \rangle$ can be enhanced as the universe evolution.  For the
$\delta < 0$ case, one may derive $z_{eff} \sim {\rm max} [4/x,
2|\delta|]$ when $|\delta| > \gamma$, we then find that $\langle
\sigma v \rangle$ has a maximum at $x \sim 2/|\delta|$. If $|\delta|
\ll \gamma$, one cannot obtain an obvious peak. When $x \gg 4/{\rm
max}[ |\delta|, \gamma]$, $\langle \sigma v \rangle$ will approach
to a constant for both cases. Our numerical results [using Eq.
(\ref{cross})] in Figs. \ref{f1} and  \ref{f2} explicitly
demonstrate the above analysis. In Fig. \ref{f1}, the parameter
$R_0$ is defined as $R_0 \equiv \langle \sigma v \rangle / \langle
\sigma v \rangle_0$, with $\langle \sigma v \rangle_{0}$ denoting
the thermally averaged annihilation cross section at $T = 0$. One
may easily see from Fig. \ref{f1} that the $\delta < 0$ case gives
the larger $R_0$ at the higher temperature. In Fig. \ref{f2}, we
plot the ratio $R_f \equiv \langle \sigma v \rangle / \langle \sigma
v \rangle_{x = 20}$. It is seen that for small $|\delta|$ and
$\gamma$ with $|\delta|\sim \gamma < 10^{-3}$, both of the cases
provide a significant enhancement for the thermally averaged
annihilation cross section. For much smaller $|\delta| = \gamma =
10^{-7}$, the $\delta < 0$ case can give the larger enhancement. It
should be mentioned that our results are independent of $\alpha$,
$B_i B_f$, and $m$.


\ssection{Boost factor and couplings} The Breit-Wigner enhancement
mechanism can change the indirect dark matter detection. On the
other hand, the Breit-Wigner enhancement will affect the calculation
of the dark matter relic density because the annihilation process
does not freeze out even after the usual ``freeze-out" time $x_f =
20$ \cite{Murayama}. More importantly, we can derive the size of
$\alpha$ ($\delta > 0$) and $B_i B_f$ ($\delta < 0$) for given
$\delta$ and $\gamma$ from the observed dark matter abundance
$\Omega_D h^2 = 0.1131 \pm 0.0034$ \cite{WMAP}. The values of
$\alpha$ and $B_i B_f$ can help us to build appropriate models. With
the help of Eqs. (\ref{neg}), (\ref{pos}) and (\ref{cross}), one can
calculate the thermally averaged annihilation cross section $\langle
\sigma v \rangle$ in the galactic halo, which allows us to obtain
the exact boost factor. It is worthwhile to stress that the
Breit-Wigner enhancement does not affect the direct dark matter
searches.

\begin{figure}[t]\begin{center}
\includegraphics[scale=0.45]{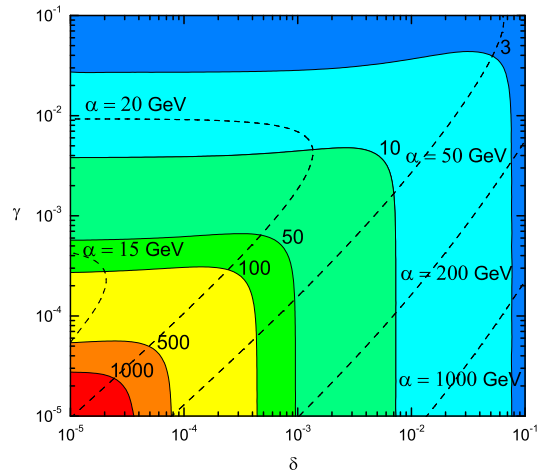}
\end{center}
\vspace{-0.6cm} \caption{Numerical illustration of the boost factor
$BF$ (solid lines) and the coupling $\alpha$ (dashed lines) on the
$\delta$ and $\gamma$ planes for the $\delta > 0$ case. }
\label{positive}\end{figure}

\begin{figure}[t]\begin{center}
\includegraphics[scale=0.45]{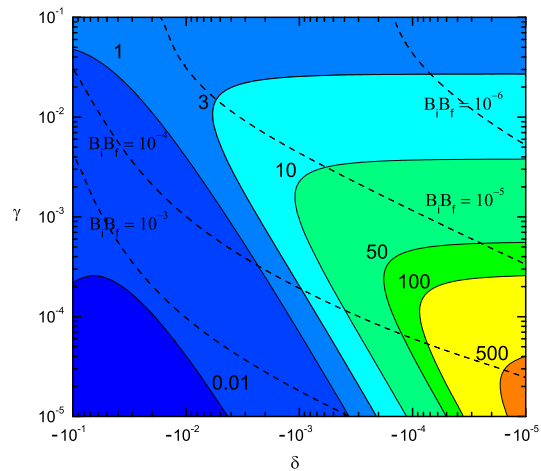}
\end{center}
\vspace{-0.6cm} \caption{Numerical illustration of the boost factor
$BF$  (solid lines) and the parameter $B_i B_f$ (dashed lines) on
the $\delta$ and $\gamma$ planes for the $\delta < 0$ case. }
\label{negative}\end{figure}

The evolution of dark matter abundance is given by the following
Boltzmann equation \cite{KOLB}:
\begin{eqnarray}
\frac{d Y}{d x} = - \frac{x \; {\bf s}(x)}{H} \langle \sigma v
\rangle (Y^2 -Y_{EQ}^2) \; , \label{bol}
\end{eqnarray}
where $Y \equiv n/{\bf s}(x)$  denotes the dark matter number
density. The entropy density ${\bf s}(x)$ and the Hubble parameter
$H$ evaluated at $x=1$ are given by
\begin{eqnarray}
{\bf s}(x) & = & \frac{2 \pi^2 g_*}{45} \frac{m^3}{x^3} \;;  \label{s}\\
H & = & \sqrt{\frac{4 \pi^3 g_*}{45}} \frac{m^2}{M_{PL}} \label{h}
\;,
\end{eqnarray}
where $M_{PL} \simeq 1.22 \times 10^{19}$ GeV is the Planck energy.
$g_*$ is the total number of effectively relativistic degrees of
freedom. Here we choose $g_* = 106.75$ for illustration. Using the
result $Y_0$ of the integration of Eq. (\ref{bol}), we may obtain
the dark matter relic density $\Omega_D h^2$
\begin{eqnarray}
\Omega_D h^2 =2.74 \times 10^8 \frac{m}{\rm GeV} Y_0 \;.
\end{eqnarray}

Using the Boltzmann equation in Eq. (\ref{bol}), we numerically
calculate $\alpha$ and $B_i B_f$ for the unphysical pole case and
the physical pole case, respectively. Our numerical results (dashed
lines) are shown in Figs. \ref{positive} and  \ref{negative}. Here
we have taken $m = 1$ TeV.  For the $\delta > 0$ case, we obtain $15
\; {\rm GeV} \lesssim \alpha \lesssim 4.9  \; {\rm TeV} $, which can
be easily satisfied. For the $\delta < 0$ case, the parameter $B_i
B_f$ is far less than the upper bound 0.25 except for the lower left
region. For most of the parameter range, $B_i B_f < 0.001$ indicates
that the successful models must have a hierarchy between the initial
branching factor $B_i$ and the final branching factor $B_f$. If the
dark matter mass $m$ is enlarged by $N$ times, $\alpha$ and $B_i
B_f$ in Figs. \ref{positive} and \ref{negative} should be
approximately enlarged by $N^2$ times.

After obtaining  $\alpha$ and $B_i B_f$, we calculate the thermally
averaged annihilation cross section $\langle \sigma v \rangle$ in
the galactic halo ($x \approx 3 \times 10^6$), which is shown in
Figs. \ref{positive} and  \ref{negative}. We would like to emphasize
that our results are insensitive to the dark matter mass $m$. Here
we normalize $\langle \sigma v \rangle$ by the usual nonresonance
annihilation cross section $10^{-9}$ ${\rm GeV}^{-2}$ to define the
boost factor $BF$
\begin{eqnarray}
BF \equiv \frac{\langle \sigma v \rangle}{10^{-9} \; {\rm GeV}^{-2}}
\;.
\end{eqnarray}
It is clear that smaller $|\delta|$ and $\gamma$ will provide larger
boost factors. For the $\delta
> 0$ case, the large boost factor ($BF \geq 100$) requires
$\delta,\gamma < {\cal O} (10^{-3})$. Our results have some
differences from  Fig. 4 in Ref. \cite{Murayama} for the region $BF
< {\cal O}(10)$ even if we choose $g_* =200$. For the $\delta < 0$
case, one may obtain $\delta,\gamma \lesssim {\cal O} (10^{-4})$ for
$BF \geq 100$. In the lower left region of Fig. \ref{negative}, we
find $BF \ll 1$, which implies that the indirect dark matter
detection experiments will not find any signal of the dark matter
annihilation. It is clear that both cases can provide a large enough
boost factor to explain the PAMELA, ATIC, and PPB-BETS anomalies.
This is one of our primary results.


\ssection{Discussion and Conclusion} We have evaluated the boost
factor $BF$ and the values of $\alpha$, $B_i B_f$ in terms of the
observed dark matter abundance $\Omega_D h^2$. If the parameter
$\alpha^2$ or $B_i B_f$ is enlarged by $N$ times, the Breit-Wigner
enhanced annihilation cross section will be enlarged by the same
times. While the dark matter relic number density $Y$ will be
approximately suppressed by $N$ times, one thus needs to introduce
new dark matter candidates. Although we can obtain the larger boost
factor, this scenario will give the smaller dark matter annihilation
rate, which is proportional to $\langle \sigma v \rangle \, Y^2$. In
fact, many models have several dominant annihilation processes,
which may include the nonresonance and resonance cases. If the
nonresonance annihilation processes determine the dark matter relic
density, one cannot obtain $\alpha$, $B_i B_f$, and $BF$ from
$\Omega_D h^2$. In this case, one may determine those parameters
from other constraints; the required values for $\alpha$, $B_i B_f$,
and $BF$ must be smaller than the predicted values in Figs.
\ref{positive} and  \ref{negative}. In general, it still has some
parameter space in this case to account for the PAMELA, ATIC, and
PPB-BETS results.

In conclusion, we have made a comprehensive analysis based on the
Breit-Wigner enhancement near the resonance point. In terms of the
observed value of $\Omega_D h^2$, we have evaluated the exact
thermally averaged annihilation cross section $\langle \sigma v
\rangle$ in the galactic halo and the boost factor $BF$, and
calculated the couplings $\alpha$ and $B_i B_f$ of the annihilation
process for both the $\delta > 0$ and the $\delta < 0$ cases,
respectively. The numerical results lead us to a general conclusion
that both the $\delta > 0$ and the $\delta < 0$ cases can provide a
large enough boost factor $BF \geq 100$ to explain the PAMELA, ATIC,
and PPB-BETS anomalies. It would be interesting to find a model with
the appropriate coupling $\alpha$ or $B_i B_f$ for the annihilation
processes to give the desired dark matter relic density.

\ssection{Acknowledgments} This work was supported by the National
Nature Science Foundation of China (NSFC) under Grants No. 10847163
and No. 10821504, and the Project of Knowledge Innovation Program
(PKIP) of the Chinese Academy of Science.


\begin{thebibliography}{99}

\bibitem{DM} For a review, see: G. Jungman, M. Kamionkowski and K.
Griest, Phys. Rept. {\bf 267}, 195 (1996); G. Bertone, D. Hooper and
J. Silk, Phys. Rept. {\bf 405}, 279 (2005).

\bibitem{WMAP} E. Komatsu {\it et al.}, arXiv:0803.0547.


\bibitem{PAMELA} O. Adriani {\it et al.},
arXiv:0810.4995; arXiv:0810.4994.

\bibitem{ATIC} J. Chang {\it et al.}, Nature {\bf 456}, 362 (2008).

\bibitem{PPB} S. Torii {\it et al.}, arXiv:0809.0760.

\bibitem{Structure} J. Lavalle, Q. Yuan, D. Maurin and X. J. Bi,
Astron. Astrophys. {\bf 479}, 427 (2008).

\bibitem{Decay}
K. Ishiwata, S. Matsumoto and T. Moroi, arXiv:0805.1133;
arXiv:0811.0250; C. R. Chen, F. Takahashi and T. Yanagida,
arXiv:0809.0792; A. E. Nelson and C. Spitzer, arXiv:0810.5167; P. F.
Yin, Q. Yuan, J. Liu, J. Zhang, X. J. Bi, S. H. Zhu and X. M. Zhang,
arXiv:0811.0176; A. Ibarra and D. Tran, arXiv:0811.1555; M. Pospelov
and M. Trott, arXiv:0812.0432; K. Hamaguchi, S. Shirai and T.
Yanagida, arXiv:0812.2374.


\bibitem{Sommerfeld} J. Hisano, S. Matsumoto, M. Nagai, O. Saito
and M. Senami, Phys. Lett. B {\bf 646}, 34 (2007); M. Cirelli, A.
Strumia, M. Tamburini, Nucl. Phys. {\bf B787}, 152 (2007); M.
Cirelli, M. Kadastik, M. Raidal and A. Strumia, arXiv:0809.2409; N.
Arkani-Hamed, D. P. Finkbeiner, T. R. Slatyer and N. Weiner,
arXiv:0810.0713; M. Lattanzi and J. Silk, arXiv:0812.0360.

\bibitem{KOLB} E. W. Kolb and M. S. Turner, {\it The Early Universe},
Addison-Wesley, Reading, MA, (1990).


\bibitem{Murayama} M. Ibe, H. Murayama and T. Yanagida,
arXiv:0812.0072.

\bibitem{Prev} M. Pospelov and A. Ritz, arXiv:0810.1502;
D. Feldman, Z. Liu and P. Nath, arXiv:0810.5762.

\bibitem{Griest} K. Griest and D. Seckel, Phys. Rev. D {\bf 43}, 3191 (1991).


\bibitem{APP}  P. Gondolo and G. Gelmini, Nucl. Phys. {\bf B360}, 145
(1991).

\bibitem{LR} W. L. Guo, L. M. Wang, Y. L. Wu, Y. F. Zhou and C.
Zhuang, arXiv:0811.2556; W. L. Guo, L. M. Wang, Y. L. Wu and C.
Zhuang, Phys. Rev. D {\bf 78}, 035015 (2008).







\end{thebibliography}
\end{document}